\documentclass[fleqn,twoside]{article}

\usepackage{espcrc2}
\usepackage{graphicx}

\newcommand{\AmS}{{\protect\the\textfont2
  A\kern-.1667em\lower.5ex\hbox{M}\kern-.125emS}}

\DeclareMathAlphabet{\mathsc}{OT1}{cmr}{m}{sc}

\newcommand{\Sol}  {\mathsc{sol}}
\newcommand{\Atm}  {\mathsc{atm}}
\newcommand{\Sbl}  {\mathsc{sbl}}
\newcommand{\Nev}  {\mathsc{nev}}
\newcommand{\Lsnd} {\mathsc{lsnd}}
\newcommand{\dms}{\Delta m^2_\Sol}
\newcommand{\dma}{\Delta m^2_\Atm}
\newcommand{\dml}{\Delta m^2_\Lsnd}

\title{
\begin{flushright} {\normalsize UWThPh-2002-29} \end{flushright}
Global analysis of neutrino oscillation data
in four-neutrino schemes}

\author{M.~Maltoni\address[ific]{Instituto de F\'{\i}sica 
  Corpuscular -- C.S.I.C./Universitat de Val{\`e}ncia \\  
  Edificio Institutos de Paterna, Apt 22085,                
  E--46071 Valencia, Spain},
T.~Schwetz\address{Institut f\"ur Theoretische Physik, Universit\"at Wien \\
  Boltzmanngasse 5, A-1090 Wien, Austria}%
  \thanks{Talk given by T.S.~at the XXX International Meeting on
  Fundamental Physics, Jaca, Spain, 28 Jan -- 1 Feb 2002. 
  This work has been supported by the ESF network 86, the  
  Marie Curie Training Grant HPMT-2000-00124 and the DOC fellowship of
  the Austrian Academy of Science.},
M.~A.~T\'ortola\addressmark[ific]
and
J.~W.~F.~Valle\addressmark[ific]}

\begin{document}

\begin{abstract}
We present an analysis of the global neutrino oscillation data in terms of
four-neutrino mass schemes. We find that the strong preference of
oscillations into active neutrinos implied by the latest solar as well as
atmospheric neutrino data allows to rule out (2+2) mass schemes,
whereas (3+1) schemes are strongly disfavoured by short-baseline
experiments. Our analysis shows that four-neutrino oscillations do not
provide a satisfactory description of the global neutrino oscillation
data including the LSND result.
\vspace{1pc}
\end{abstract}

\maketitle

\section{INTRODUCTION}
The neutrino oscillation interpretations of the solar~\cite{solar,sno}
and atmospheric~\cite{skatm,macro} neutrino data and the LSND
experiment~\cite{lsnd} require three neutrino mass-squared differences
of different orders of magnitude. Since it is not possible to obtain
this within the Standard Model framework of three active neutrinos it
has been proposed to introduce a light sterile neutrino~\cite{sterile}
to reconcile all the experimental hints for neutrino
oscillations. Here we present an analysis of the global neutrino
oscillation data in terms of four-neutrino mass schemes, including
data from solar and atmospheric neutrino experiments, the LSND
experiment, as well as data from short-baseline (SBL) experiments
\cite{KARMEN,CDHS,bugey} and long-baseline reactor experiments
\cite{lbl} reporting no evidence for oscillations. This analysis
updates the work presented at the meeting and is based on the data of
summer 2002. We find that for all possible types of four-neutrino
schemes different sub-sets of the data are in serious disagreement and
hence, four-neutrino oscillations \textit{do not} provide a
satisfactory description of the global oscillation data including
LSND. The details of our calculations can be found in
Refs.~\cite{4nu01,4nu02,solat02}.

\begin{figure}[t]
 \centering
   \includegraphics[width=0.95\linewidth]{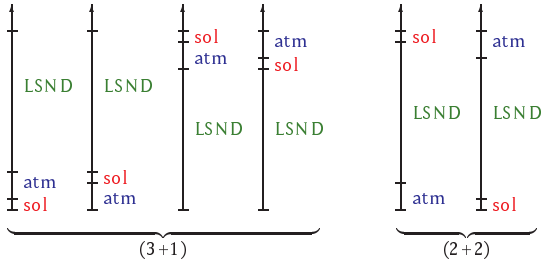}
      \caption{\label{fig:4spectra}%
        The six types of four-neutrino mass spectra, divided into the
   classes (3+1) and (2+2).}
\end{figure}

Four-neutrino mass schemes are usually divided into
the two classes (3+1) and (2+2), as illustrated in Fig.~\ref{fig:4spectra}. 
We note that (3+1) mass spectra include the three-active
neutrino scenario as limiting case. In this case solar and atmospheric
neutrino oscillations are explained by active neutrino oscillations,
with mass-squared differences $\dms$ and $\dma$, and the fourth
neutrino state gets completely decoupled. We will refer to such
limiting scenario as (3+0). In contrast, the (2+2) spectrum is
intrinsically different, as there must be a significant contribution
of the sterile neutrino either in solar or in atmospheric neutrino
oscillations or in both.

\section{NOTATIONS AND APPROXIMATIONS}
Neglecting CP violation, in general neutrino oscillations in
four-neutrino schemes are described by 9 parameters: 3 mass-squared
differences and 6 mixing angles in the unitary lepton mixing matrix.
Here we use a parameterization introduced in Ref.~\cite{4nu01}, which
is based on physically relevant quantities: the 6 parameters $\dms$,
$\theta_\Sol$, $\dma$, $\theta_\Atm$, $\dml$, $\theta_\Lsnd$ are
similar to the two-neutrino mass-squared differences and mixing angles
and are directly related to the oscillations in solar, atmospheric and
the LSND experiments. For the remaining 3 parameters we use
$\eta_s,\eta_e$ and $d_\mu$. Here, $\eta_s \,(\eta_e)$ is the fraction
of $\nu_s \,(\nu_e)$ participating in solar oscillations, and
($1-d_\mu$) is the fraction of $\nu_\mu$ participating in oscillations
with $\dma$ (for exact definitions see Ref.~\cite{4nu01}). For the
analysis we adopt the following approximations:
\begin{itemize}
\item 
We make use of the hierarchy
\begin{equation} 
\dms \ll \dma \ll \dml \,.
\end{equation}
This means that for each data set we consider only one mass-squared
difference, the other two are set either to zero or to infinity.
\item
In the analyses of solar and atmospheric data (but not for SBL data) we
set $\eta_e = 1$, which is justified because of strong constraints
from reactor experiments~\cite{bugey,lbl}.
\end{itemize}

\begin{figure}[t]
 \centering
   \includegraphics[width=0.95\linewidth]{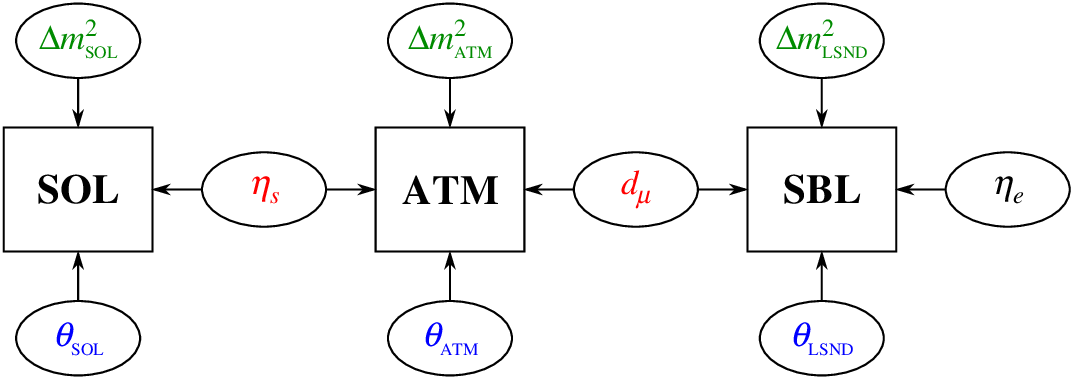}
      \caption{\label{fig:diagram}%
         Parameter dependence of the different data sets in our
         parameterization. }
\end{figure}

Due to these approximations the parameter structure of the
four-neutrino analysis gets rather simple. The parameter dependence of
the different data sets solar, atmospheric and SBL is illustrated in
Fig.~\ref{fig:diagram}. We see that only $\eta_s$ links solar and
atmospheric data and $d_\mu$ links atmospheric and SBL data.  All the
other parameters are ``private'' to one data set.

\section{(2+2): RULED OUT BY SOLAR AND ATMOSPHERIC DATA}
\label{sec:2+2}
The strong preference of oscillations into active neutrinos in solar
and atmospheric oscillations leads to a direct conflict in (2+2)
oscillation schemes. We will now show that thanks to the new SNO solar
neutrino data \cite{sno} and the improved SK statistic on atmospheric
neutrinos \cite{skatm} the tension in the data has become so strong
that (2+2) oscillation schemes are essentially ruled
out.\footnote{Details of our analyses of the solar and atmospheric
neutrino data can be found in Ref.~\cite{solat02}. For an earlier
four-neutrino analysis of solar and atmospheric data see
Ref.~\cite{concha4nu}.}  Indeed, latest solar neutrino data lead to
the bound $\eta_s \le 0.45$ at 99\% CL, where $\eta_s$ is the
parameter describing the fraction of the sterile neutrino
participating in solar neutrino oscillations. In contrast, in (2+2)
schemes atmospheric data imply $\eta_s \ge 0.65$ at 99\% CL, in clear
disagreement with the bound from solar data.

\begin{figure}[t]
  \centering
  \includegraphics[width=0.95\linewidth]{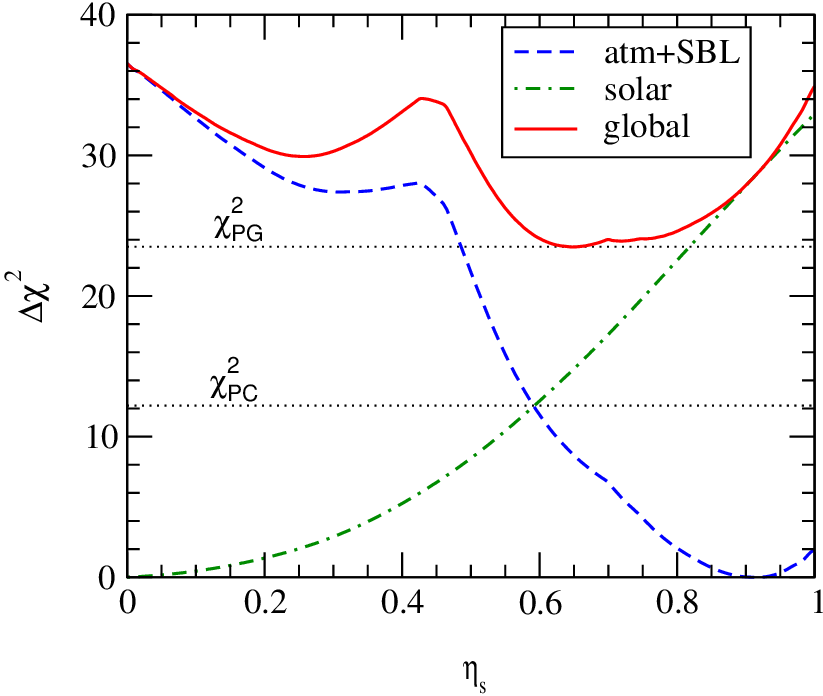}
  \caption{$\Delta\chi^2_\Sol$, $\Delta\chi^2_{\Atm+\Sbl}$
    and $\bar\chi^2_\mathrm{global}$ as a function of $\eta_s$ in
    (2+2) oscillation schemes.}
  \label{fig:sol+atm}
\end{figure}

In Fig.~\ref{fig:sol+atm} we show the $\chi^2$ for solar data and for
atmospheric combined with SBL data as a function of $\eta_s$.
Furthermore, we show the $\chi^2$ of the global data defined by
\begin{equation}\label{chi2solatm}
\bar\chi^2(\eta_s) \equiv 
\Delta\chi^2_\Sol(\eta_s) + 
\Delta\chi^2_{\Atm + \Sbl}(\eta_s) \,.
\end{equation}
From the figure we find that only if we take both data sets at the
99.95\% CL a value of $\eta_s$ exists, which is contained in the
allowed regions of both sets. This follows from the $\chi^2$-value
$\chi^2_\mathrm{PC} = 12.2$ shown in the figure. In Ref.~\cite{4nu02}
we have proposed a statistical method to evaluate the disagreement of
different data sets in global analyses. The \textit{parameter
goodness of fit} (PG) makes use of the $\bar\chi^2$ defined in
Eq.~(\ref{chi2solatm}). This criterion evaluates the GOF of the
\textit{combination} of data sets, without being diluted by a large
number of data points, as it happens for the usual GOF criterion (for
details see Ref.~\cite{4nu02}). We find $\chi^2_\mathrm{PG} \equiv
\bar\chi^2_\mathrm{min} = 23.5$, leading to the marginal PG
of $1.3 \times 10^{-6}$.  We conclude that (2+2) oscillation schemes
are highly disfavoured by the disagreement between the latest solar
and atmospheric neutrino data. This is a very robust result,
independent of whether LSND is confirmed or disproved.

\section{(3+1): STRONGLY DISFAVOURED BY SBL DATA}

\begin{figure}[t]
  \centering 
  \includegraphics[width=0.95\linewidth]{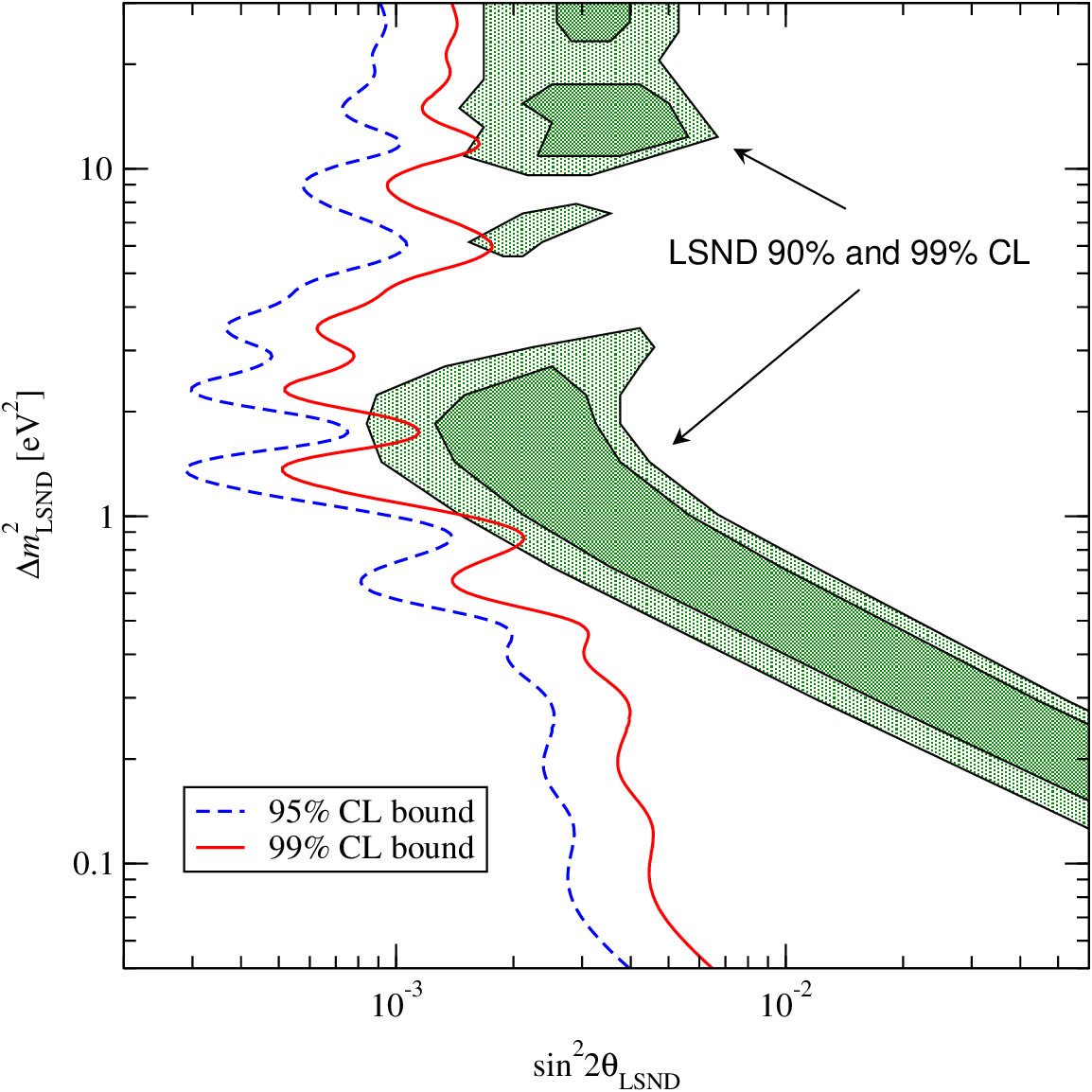}
  \caption{Upper bound on $\sin^22\theta_\Lsnd$ from SBL and
  atmospheric neutrino data in (3+1) schemes \cite{cornering} compared
  to the LSND allowed region \cite{lsnd}.}
  \label{fig:3+1}
\end{figure}

It is known for a long time \cite{3+1early} that (3+1) mass schemes
are disfavoured by the comparison of SBL disappearance data
\cite{CDHS,bugey} with the LSND result. In Ref.~\cite{cornering} we
have calculated an upper bound on the LSND oscillation amplitude
$\sin^22\theta_\Lsnd$ resulting from SBL and atmospheric neutrino
data. From Fig.~\ref{fig:3+1} we see that this bound is incompatible
with the signal observed in LSND at the 95\% CL. Only marginal overlap
regions exist between the bound and LSND if both are taken at 99\%
CL. An analysis in terms of the parameter goodness of fit \cite{4nu02}
shows that for most values of $\dml$ SBL and atmospheric data are
compatible with LSND only at more than $3\sigma$, with one exception
around $\dml \sim 6$ eV$^2$, where the PG reaches 1\%. These results
show that (3+1) schemes are strongly disfavoured by SBL disappearance
data.

\section{COMPARING (3+1), (2+2) AND (3+0) HYPOTHESES}

With the methods developed in Ref.~\cite{4nu01} we are able to perform
a global fit to the oscillation data in the four-neutrino
framework. This approach allows to statistically compare the different
hypotheses. Let us first evaluate the GOF of (3+1) and (2+2) spectra
with the help of the PG method described in Ref.~\cite{4nu02}.  We
divide the global oscillation data in the 4 data sets SOL, ATM, LSND
and NEV, where NEV contains the experiments KARMEN \cite{KARMEN}, CDHS
\cite{CDHS}, Bugey \cite{bugey} and CHOOZ/Palo Verde \cite{lbl},
reporting no evidence for oscillations. Following Ref.~\cite{4nu02}
we consider
\begin{equation}\label{chi2bar}
\begin{array}{ccl}
\bar\chi^2 &=&
\Delta\chi^2_\Sol(\theta_\Sol,\dms,\eta_s) \\
&+& \Delta\chi^2_\Atm(\theta_\Atm,\dma,\eta_s,d_\mu) \\
&+& \Delta\chi^2_\Nev(\theta_\Lsnd,\dml,d_\mu,\eta_e) \\
&+& \Delta\chi^2_\Lsnd(\theta_\Lsnd,\dml) \,,  
\end{array}
\end{equation}
where $\Delta\chi^2_X = \chi^2_X - (\chi^2_X)_\mathrm{min}$ ($X$ =
SOL, ATM, NEV, LSND). In Tab.~\ref{tab:pg} we show the contributions of
the 4 data sets to $\chi^2_\mathrm{PG} \equiv \bar\chi^2_\mathrm{min}$
for (3+1) and (2+2) oscillation schemes. As expected we observe that
in (3+1) schemes the main contribution comes from SBL data due to the
tension between LSND and NEV data in these schemes. For (2+2)
oscillation schemes a large part of $\chi^2_\mathrm{PG}$ comes from
solar and atmospheric data, however, also SBL data contributes
significantly.  This comes mainly from the tension between LSND and
KARMEN \cite{Church:2002tc}, which does not depend on the mass scheme
and, hence, also contributes in the case of (2+2). Therefore, the
values of $\chi^2_\mathrm{PG}$ in Tab.~\ref{tab:pg} for (2+2) schemes
are higher than the one given in Sec.~\ref{sec:2+2}, where the tension
in SBL data is not included.

\begin{table}[t]\centering
    \catcode`?=\active \def?{\hphantom{0}}
    \begin{tabular}{|c|cccc|c|}
    \hline
    & SOL & ATM & LSND & NEV &   $\chi^2_\mathrm{PG}$ \\
    \hline 
(3+1) & ?0.0 & 0.4 & 7.2 & 7.0 & 14.6  \\
(2+2) & 14.8 & 6.7 & 2.2 & 9.7 & 32.4  \\
    \hline
    \end{tabular}
    \caption{\label{tab:pg}%
Contributions of different data sets to $\chi^2_\mathrm{PG}$ 
in (3+1) and (2+2) schemes.}
\end{table}

The parameter goodness of fit is now obtained by evaluating
$\chi^2_\mathrm{PG}$ for 4 DOF \cite{4nu02}. This number of degrees of
freedom corresponds to the 4 parameters $\eta_s, d_\mu, \theta_\Lsnd, \dml$
describing the coupling of the different data sets (see
Eq.~(\ref{chi2bar})). The best GOF is obtained in the (3+1)
case. However, even in this best case the PG is only 0.56\%. The PG of
$1.6\times 10^{-6}$ for (2+2) schemes shows that these mass schemes
are essentially ruled out by the disagreement between the individual
data sets.

Although we have seen that none of the four-neutrino mass schemes can
provide a reasonable good fit to the global oscillation data including
LSND, it might be interesting to consider the \textit{relative} status
of the three hypotheses (3+1), (2+2) and the three-active neutrino
scenario (3+0). This can be done by comparing the $\chi^2$ values of
the best fit point (which is in the (3+1) scheme) with the one
corresponding to (2+2) and (3+0).  First we observe that (2+2) schemes
are strongly disfavoured with respect to (3+1) with a $\Delta \chi^2 =
17.8$. The reason for the big change with respect to the value of
$\Delta \chi^2 = 3.7$ found in Ref.~\cite{4nu01} is the improved
sensitivity of solar (SNO NC) and atmospheric (SK 1489-days) data to a
sterile component. With this new data now (3+1) schemes are clearly
preferred over (2+2): for 4 DOF a $\Delta \chi^2 = 17.8$ implies that
(2+2) is ruled out at 99.87\% CL with respect to (3+1). Further we
find that (2+2) is only slightly better than (3+0), which is
disfavoured with a $\Delta \chi^2 = 20.0$ with respect to (3+1).

\section{CONCLUSIONS}
Using latest solar and atmospheric data, we re-analyze the
  four-neutrino description of current global neutrino oscillation
  data, including the LSND evidence for oscillations.  The higher
  degree of rejection for non-active solar and atmospheric oscillation
  solutions implied by the SNO neutral current result as well as by
  the latest 1489-day SK atmospheric neutrino data allows us to rule
  out (2+2) oscillation schemes. Using an improved goodness of fit
  method especially sensitive to the combination of data sets we
  obtain a GOF of only $1.6\times 10^{-6}$ for (2+2) schemes. Further,
  we find that also (3+1) schemes are strongly disfavoured by the
  data. The disagreement between negative short-baseline experiments
  and LSND in (3+1) schemes imply a GOF of $5.6\times 10^{-3}$.
  This leads to the conclusion that all four-neutrino descriptions of
  the LSND anomaly, both in (2+2) as well as (3+1) realizations, are
  highly disfavoured.  Our analysis brings the LSND hint to a more
  puzzling status, and the situation will become even more puzzling if
  LSND should be confirmed by the up-coming MiniBooNE experiment
  \cite{miniboone}.

\end{document}